\documentclass[twocolumn,aps,preprintnumbers,superscriptaddress]{revtex4}
\usepackage{graphicx}
\usepackage{bm}
\usepackage{amsmath, amsthm, amssymb}

\begin{document}
\title{ Degeneracy of cross sections in scattering of light}
\author{Jeng Yi Lee}
\affiliation{
Department of Opto-Electronic Engineering, National Dong Hwa University, Hualien 974301, Taiwan }

\date{\today}

\begin{abstract} 
We theoretically and numerically prove that under an electromagnetic plane wave with linear polarization incident normally to cylindrical passive scatterers, a single energy diagram can integrate absorption, scattering, and extinction cross sections for arbitrary scattering systems, irrespective of internal configuration, material parameters, and its sizes. 
For a system with definite resonant orders, along the corresponding boundary of the energy diagram, not only the magnitudes of scattering coefficients, but also its phases, are required the same, corresponding to superabsorption or superscattering.
For systems composed by larger resonant orders, the domain of the energy diagram can completely cover systems with lower one.
Hence, systems with different resonant orders can provide the same energy characteristics, reflecting the energy degenerate property in light scattering.
This degeneracy may relax more degrees of freedom in functional designs, such as energy harvesting, imaging, sensing devices.
We demonstrate various systems  based on real materials to support this finding.
\end{abstract}
\pacs{ }

\maketitle

Understanding of the limitation of interaction between scatterers and  electromagnetic radiation is of importance for a variety of applications, such as solar energy harvesting \cite{energy1,energy2,energy3}, strengthening the interaction of light and subwavelength objects \cite{enhance1}, bio-imaging \cite{imaging1,imaging2}, highly directivity of antennas \cite{super1,kerker1,kerker2}, sensing \cite{sensor}, and medical heating treatment at nanoscales \cite{heating1,heating2,heating3}.

Scattering and absorption of light by an extremely subwavelength scatterer with an isotropic and homogeneous structure can not go beyond a single resonant order \cite{book1,single1,single2}.
To overcome this limit, a composite subwavelength system by proper multi-layered configurations can induce multiple resonances.
It has been demonstrated that, in the whispering gallery condition, excitation of a confined polariton can lead to a system with a degenerate resonant super-scattering \cite{superscattering1,superscattering2,superscattering3}.
Analogous to degenerate resonant mechanism, superabsorber, that not only can retain minimum scattering power but also can arbitrarily strengthen absorption, was demonstrated, \cite{superabsorber}.

We note that due to passive materials embedded, physical energy conservation law must be met, that is often overlooked but important.
By this physical law, the corresponding partial absorption, scattering, and extinction cross sections are definitely bounded, irrespective of inherent system configurations, material parameters, and operating environment \cite{phase1,phase2}.
Consequently, as a benefit of the obtained phase diagram for scattering coefficients, we can find that for each resonant order, scatterer systems can support a variety of energy absorption with constant scattering power.
However, when encountering multiple resonant orders interference, the answer to whether there can have an energy diagram to indicate arbitrary scattering configurations is far from obvious and remains open.
The answer can definitely provide global information in light scattering and also understand the inherent limits, with potential applications in the development of nano-photonics devices.

Inspired by the concept from the calculus of variations \cite{landau}, in this work, we borrow differential calculus as well as the employed Lagrange multiplier to address energy distribution for scattering of light.
We theoretically and numerically prove that under an electromagnetic plane wave with linear polarization excitation,  for a cylindrical system, there exists one energy diagram where can integrate all possible scattering configurations.
The energy diagram can reveal an one-to-one correspondence for absorption, scattering, and extinction cross sections.
But, on the contrary, there has no such one-to-one relation for scattering coefficients, reflecting the degenerate property in light scattering.
For a system with definite resonant orders,  along the corresponding domain boundary, the magnitudes of scattering coefficients and its phase are required, which corresponds to superabsorption or superscattering.
In addition, we observe that for the scattering system with lower scattering orders, its domain is just the subspace of higher ones.
Therefore, one can design a scatterer system, through exciting high orders, to simulate the identical energy performance occurred at low orders.
We demonstrate quasi-superscattering and quasi-superabsorption to support this finding.
We believe our results can provide more degrees of freedom for nano-photonics designs in energy harvesting, imaging, sensing.

Consider a cylindrically symmetric scatterer is normally  impinged by a plane wave with electric field oscillated along the z-axis direction, i.e., \textbf{s} mode.
The scattering, extinction, and absorption powers can be expressed in the following \cite{book1,superscattering1,linear},
\begin{equation}
\begin{split}
P_{ext}&=P_{scat}+P_{abs}=-\frac{2}{k_0}\sqrt{\frac{\epsilon_0}{\mu_0}}\vert E_0\vert^2\sum_{n=-\infty}^{\infty}\text{Re}(a_n^{\textbf{s}})\\
P_{abs}&=-\frac{2}{k_0}\sqrt{\frac{\epsilon_0}{\mu_0}}\vert E_0\vert^2\sum_{n=-\infty}^{\infty}[\text{Re}(a_n^{\textbf{s}})+\vert a_n^{\textbf{s}} \vert^2]\\
\end{split}
\end{equation}
where $E_0$ is amplitude of incident electric field, $a_n^{\textbf{s}}$ is complex scattering coefficient, $k_0$ is environmental wave number, and $\epsilon_0$ and $\mu_0$ are free space permittivity and permeability \cite{note}.
Without loss of generality we set  $\vert E_0\vert=1$.
We note that due to cylindrical symmetry, the scattering coefficients have a symmetry for $a_n^{\textbf{s}}=a_{-n}^{\textbf{s}}$.
For $n=[0,1,2]$, they correspond to electric dipole, magnetic dipole, and magnetic quadrupole, respectively.
The extinction power, summation of absorption and scattering power, is also related to the optical theorem, which links the  scattering electric field at the forward direction \cite{optical,optical1,optical2,optical3}.
Thus, systems with non-zeros of energy dissipation and radiation, they must possess a non-zero electric field along the forward direction.
The promotion of absorption, under constant extinction, would reduce the scattering, that would further lower the scattering intensity over all direction, but it still remains the constant for  scattering electric field in the forward direction.
Now, due to passive material embedded, the  partial absorption cross sections for each orders  would be restricted $-[\text{Re}(a_n^{\textbf{s}})+\vert a_n^{\textbf{s}}\vert^2]\geq 0$, because of energy conservation.
Here $-[\text{Re}(a_n^{\textbf{s}})+\vert a_n^{\textbf{s}}\vert^2]$ is defined as normalized partial absorption power.
Following this inequality, we obtain a clear physical bound on the scattering coefficients for orders, i.e., $\vert a_n^{\textbf{s}}\vert\in[0,1]$ and $Arg[a_n^{\textbf{s}}]\in[\frac{\pi}{2},\frac{3\pi}{2}]$ \cite{kerker1,phase1}.

For each order, the normalized partial absorption power can not go beyond $0.25$, while the upper limit for normalized partial  scattering power, defined as $\vert a_n^{\textbf{s}}\vert^2$, is $1$.
Therefore, if a system has a $N$ orders excited, the maximum absorption power can not be larger than $(2N+1)\times \frac{1}{2k_0}\sqrt{\frac{\epsilon_0}{\mu_0}}$ while the corresponding scattering power would be $(2N+1)\times \frac{1}{2k_0}\sqrt{\frac{\epsilon_0}{\mu_0}}$, corresponding coherent perfect absorption \cite{absorption}. 
In an extreme situation, the maximum scattering power in $N$ orders can reach $(2N+1)\times \frac{2}{k_0}\sqrt{\frac{\epsilon_0}{\mu_0}}$, while its corresponding absorption is zero.
Then, it is naturally extended to seek for the existence of the energy diagram when more complicated resonance orders excited.

\begin{figure}[b]
\centering
\includegraphics[width=9cm]{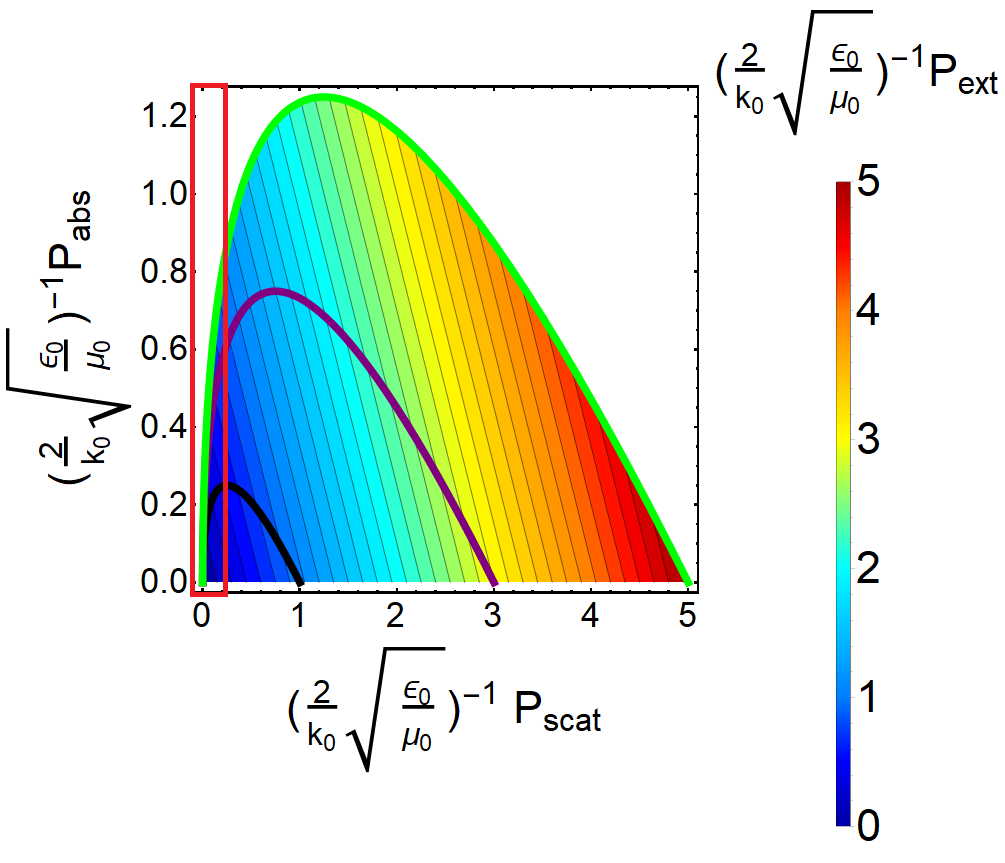}
\caption{(Color online) Integrated energy diagram for normalized absorption, scattering, and extinction powers, applicable to general scattering states.  The black line indicates the limit boundary  for a system with $N=0$ order dominant, the purple is for a system with $N=[-1,0,1]$ orders excited, and the green is for $N=[-2,-1,0,1,2]$ orders excited. The red line box highlights a region with a lower scattering, but with an arbitrary absorption enhanced when more resonant orders are involved. Note that each domain is a subspace of the higher dimensional scattering system.} 
\end{figure} 

To address such a fundamental problem, we use the differential calculus as well as the Lagrange multiplier to figure out the energy diagram under given dissipated power \cite{math,landau}.
Suppose our system has a fixed dissipation power $P_{abs}=Const.$, we expect to evaluate the extreme scattering power $P_{scat}$.
Now, $P_{abs}=Const.$ is a constraint with primary $a_{-N}^{\textbf{s}}$ to $a_{N}^{\textbf{s}}$ involved, i.e.,
\begin{equation}
\begin{split}
&P_{abs}(a_{-N}^{\textbf{s}},..0,..a_{-N}^{\textbf{s}})\\
&=-\frac{2}{k_0}\sqrt{\frac{\epsilon_0}{\mu_0}}\sum_{n=-N}^{N}[\text{Re}(a_n^{\textbf{s}})+\vert a_n^{\textbf{s}} \vert^2]=const.
\end{split}
\end{equation}
Here we have no assumption for the scattering coefficients.

Then, we define a energy function $\textit{L}$ related to scattering and absorption powers as follows
\begin{equation}
\begin{split}
&\textit{L}(a_{-N}^{\textbf{s}},..,0,..a_{N}^{\textbf{s}})=
P_{scat}(a_{-N}^{\textbf{s}},..0,..a_{-N}^{\textbf{s}})\\
&+\lambda  P_{abs}(a_{-N}^{\textbf{s}},..0,..a_{-N}^{\textbf{s}})\\
&=\frac{2}{k_0}\sqrt{\frac{\epsilon_0}{\mu_0}}\sum_{n=-N}^{N}[\vert a_n^{\textbf{s}}\vert^2-\lambda \vert a_n^{\textbf{s}}\vert\cos\theta_n-\lambda \vert a_n^{\textbf{s}}\vert^2].
\end{split}
\end{equation}
where $\lambda$ is a Lagrange multiplier, and we have used the phasor representation for complex scattering coefficients, $a_n^{\textbf{s}}=\vert a_n^{TE}\vert e^{i\theta_n}$ here $\theta_n$ is the argument of $a_n^{\textbf{s}}$.

To have the extreme minimum or maximum of the scattering powers, 
the energy function $\textit{L}$  is required to meet,
\begin{equation}\label{condition}
\begin{split}
&\frac{\partial \textit{L}}{\partial \vert a_n^{\textbf{s}}\vert}\\
&=\frac{2}{k_0}\sqrt{\frac{\epsilon_0}{\mu_0}}[2\vert a_n^{\textbf{s}}\vert-\lambda \cos\theta_n-2\lambda\vert a_n^{\textbf{s}}\vert]=0,\\
&\frac{\partial \textit{L}}{\partial \theta_n}\\
&=\frac{2}{k_0}\sqrt{\frac{\epsilon_0}{\mu_0}}[\lambda \vert a_n^{\textbf{s}}\vert\sin\theta_n]=0,
\end{split}
\end{equation}
which are valid for $n=-N$ to $N$.
In the latter expression,  to obtain non-trivial solutions for $\lambda$ and $a_n^{\textbf{s}}$, we obtain $\theta_n=0$ or $\theta=\pi$.
However, by energy conservation, the only applicable solution   would be $\theta_n=\pi$.
By this outcome, the first expression in Eq.(\ref{condition}) would be
\begin{equation}
\begin{split}
2\vert a_n^{\textbf{s}}\vert +\lambda-2\lambda \vert a_n^{\textbf{s}}\vert =0.
\end{split}
\end{equation}
Thus, we have $\vert a_n^{\textbf{s}}\vert =\frac{\lambda}{2\lambda-2}\equiv s.$
Further, $a_n^{\textbf{s}}=-s$, that the magnitudes of scattering coefficients are identical and the corresponding phases are $\pi$, for each orders.

Consequently, we can express the scattering coefficients in terms of absorption power,
\begin{equation}
\begin{split}
a_n^{\textbf{s}}&=\frac{-1\pm\sqrt{1-\frac{2P_{abs}k_0}{(2N+1)}\sqrt{\frac{\mu_0}{\epsilon_0}}}}{2}
\end{split}
\end{equation}
that there are two extrinsic scattering performances under given absorption.
 
In the square root, it can guarantee the real value, because $P_{abs}\leq \frac{2N+1}{2k_0}\sqrt{\frac{\epsilon_0}{\mu_0}}$ for N orders dominant.
The ultimate limit would be maxima absorption, i.e., $\frac{2N+1}{2k_0}\sqrt{\frac{\epsilon_0}{\mu_0}}$, corresponding to coherent perfect absorption \cite{absorption}.
More appealingly, our results imply that under constant absorption power, there can support two solutions for scattering powers:
\begin{equation}
\begin{split}
P_{scat}^{Max}&=\frac{1}{k_0}\sqrt{\frac{\epsilon_0}{\mu_0}}(2N+1)[1+\sqrt{1-\frac{2P_{abs}k_0}{(2N+1)}\sqrt{\frac{\mu_0}{\epsilon_0}}}\\
&-\frac{P_{abs}k_0}{(2N+1)}\sqrt{\frac{\mu_0}{\epsilon_0}}],\\
P_{scat}^{Min}&=\frac{1}{k_0}\sqrt{\frac{\epsilon_0}{\mu_0}}(2N+1)[1-\sqrt{1-\frac{2P_{abs}k_0}{(2N+1)}\sqrt{\frac{\mu_0}{\epsilon_0}}}\\
&-\frac{P_{abs}k_0}{(2N+1)}\sqrt{\frac{\mu_0}{\epsilon_0}}].\\
\end{split}
\end{equation}

\begin{figure*}
\includegraphics[width=18cm]{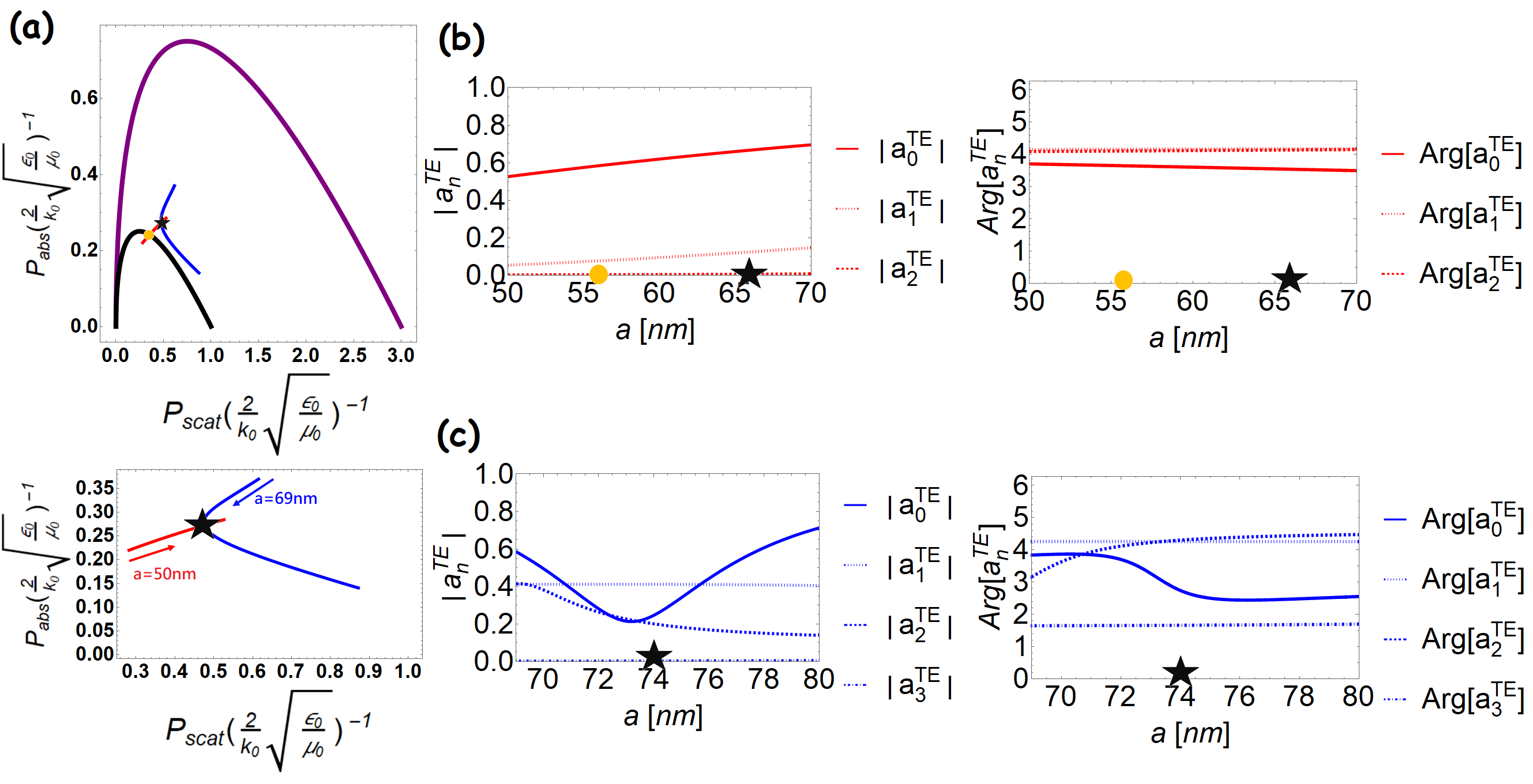}
\caption{(Color online)(a) Energy distribution evolutions of two different configuration systems  by tuning the geometry sizes. Red line is from a core-shell system made by gold in shell and silicon in core. Here the radius ratio by this core-shell (inner to outer) is fixed constant, $0.2$. The blue is from a homogeneous silicon system. We can see two systems have the same energy distribution in absorption, scattering, and extinction, highlight by a black star. The inset figure (below) enlarges its energy distribution details. The magnitudes and  phases of scattering coefficients for each orders for the gold-shell and silicon-core system by tuning geometric sizes are shown in Left and Right sides of (b). 
The relative permittivity parameter of gold is $-2.81 + i3.19$ and that of silicon is $18.5+ i0.63$. In (c), we analyze the magnitudes and phases from a silicon-based system with the $n=0,1,2,3$, respectively. } 
\end{figure*}

Now, we depict an energy diagram involving normalized scattering, absorption, and extinction cross sections as shown in Fig. 1 \cite{note2}.
The boundary of the energy diagram is followed by $[(\frac{2}{k_0}\sqrt{\frac{\epsilon_0}{\mu_0}})^{-1}P_{abs},(\frac{2}{k_0}\sqrt{\frac{\epsilon_0}{\mu_0}})^{-1}P_{scat}]=[(s-s^2)(2N+1),s^2(2N+1)]$ where $s\in [0,1]$ for $N$ resonance orders dominant.
We construct the density plot for normalized extinction power, defined  as $(\frac{2}{k_0}\sqrt{\frac{\epsilon_0}{\mu_0}})^{-1}P_{ext}$, in Fig. 1.
It clearly reveals the limit of normalized absorption, scattering, and extinction powers.
Here $N=0$ represents a system with only electric dipole resonance supported, $N=1$ has electric and magnetic dipoles resonances, while a system with $N=2$ consists of electric, magnetic dipoles, and magnetic quadrupole resonances.
The maximum normalized extinction power accompanies with the maxima of normalized scattering power, while the corresponding normalized absorption is zero.
Additionally, with the constant normalized extinction power, the  increasing normalized absorption (dissipation) can reduce the output scattering performance.
We note that by considering optical theorem, this constant normalized extinction would require the fixed cost in forward electric field, but it could have a reduction in scattering distribution over all direction when absorption involved.
Interestingly, within the regime of lower scattering, as indicated by red line box in Fig. 1, the energy diagram reflects the existence of a higher absorption power when multiple resonant orders are properly involved, corresponding to a superabsorption \cite{superabsorber}.
We will discuss this interesting scattering property in the following.
Moreover, although for N resonant orders, it has a definite domain boundary, the corresponding domain is just a subspace of a higher orders system, which implies that the scattering states  are degenerate in energy cross sections.
\begin{figure*}
\includegraphics[width=18cm]{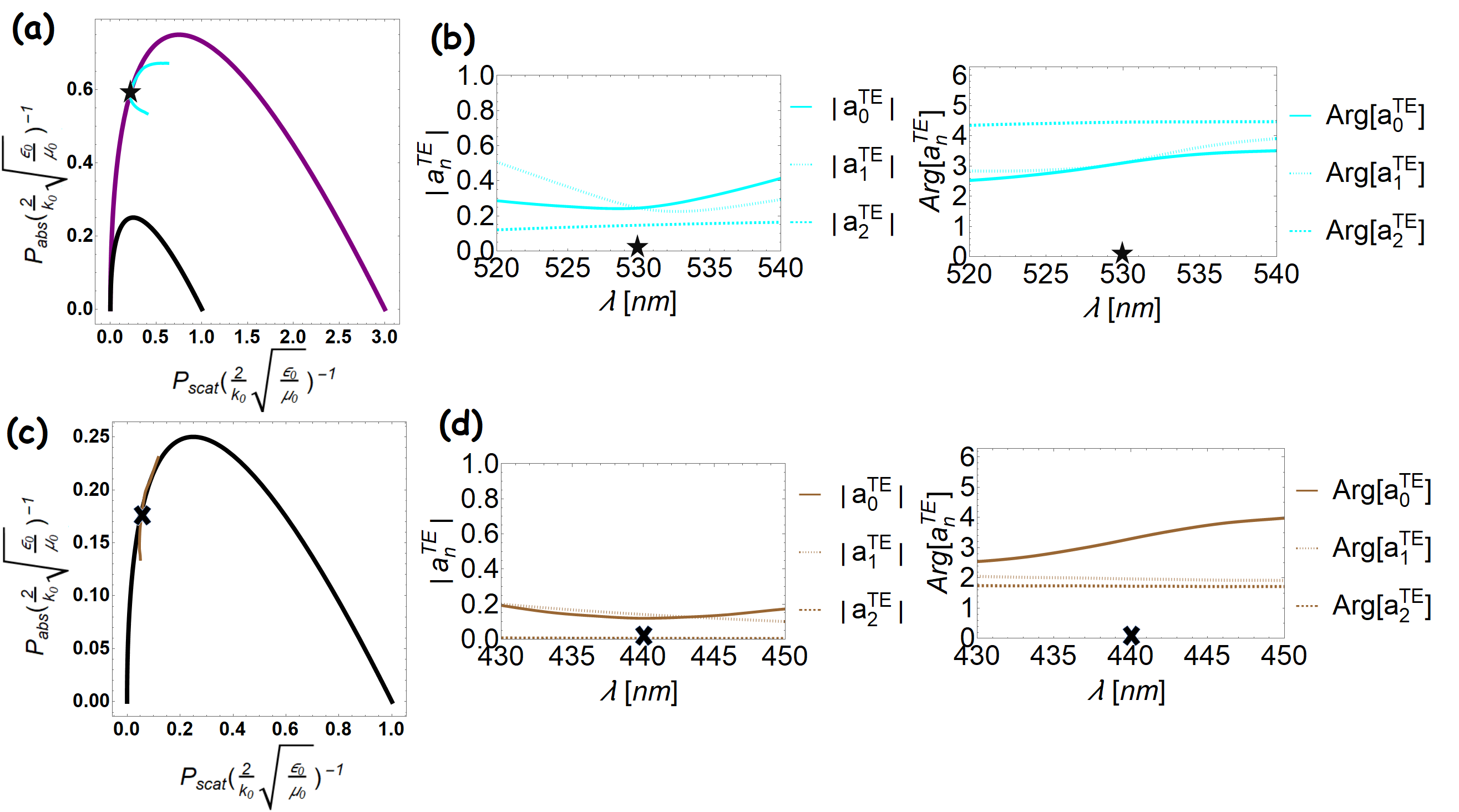}
\caption{(Color online) (a) Evolutions of energy distribution  of two different configuration systems  by tuning the operating wavelength. The cyan color line represents the silicon-shell and gold-core system with an outer radius of $132.98$nm and a ratio of the inner to outer radius of $0.42$. The magnitudes and phases of scattering components are shown in (b).  There is a single intersection at $N=1$ boundary at wavelength $530$nm, marked by a black star. In (c), we consider another core-shell system with silicon in shell and silver in core with outer radius being $51.6$nm and ratio of outer-inner radius being $0.89$. The evolution of energy distribution is from wavelength from $430$nm to $450nm$ shown in a brown color line. There has a intersection point in $N=0$ boundary, marked by a black cross. The magnitudes and phases of scattering components by this wavelength range are shown in (d). A marked cross denotes the intersection point is at $440$nm.} 
\end{figure*}
As a result, the scattering system can display the same energy performance, but its inherent constitution is completely different.
We emphasize that only the boundary for the corresponding resonant system requires a definite constraint in scattering coefficients, but there have no such one-to-one correspondences for that inside the domain.

To verify our finding, in Fig. 2, we first discuss the systems with realistic materials embedded \cite{handbook}.
 We choose two systems with different geometrical sizes, materials, and inherent configurations, but at the same operating wavelength $500$nm.
In Fig. 2 (a), the result by a red line represents that a core-shell nanowire system is constituted by golden in shell and silicon in core, while the blue line is from a homogeneous silicon nanowire.
We tune the outer radius $a$ from $50$nm to $70$nm for the core-shell system, while the ratio of shell-core radii is fixed constant $0.2$.
For the homogeneous silicon nanowire, we tune the radius from $69$nm to $80$nm.
Initially, the rad line residues in $N=0$ region when the outer radius is $50$nm, its energy distribution then goes through $N=1$ region when the outer radius is larger than $55$nm marked by a yellow point in Fig. 2 (a).
In order to understand its underlying mechanism, we investigate the magnitudes and phases of the scattering coefficients for each orders in Fig. 2 (b).
We can observe that the dominant order is by $n=0$ at   $a=50$nm, however, with the increasing geometrical size, the $n=1$ order would gradually become another primary contribution.
We also find that the phase from the primarily contributed order is nearly $\pi$ when close to the boundary with $N=0$, as expected.

For the system made by only silicon, we depict a blue line for a relation between the normalized absorption and scattering powers by geometry size from $69$nm to $80$nm in Fig. 2 (a).
We can see, in Fig. 2 (c), that the contributed orders are $n=[-2,-1,0,1,2]$, which obviously belong to scattering space $N=2$.
This outcome implies that the domain should be $N=2$, but its location also crosses over $N=1$, which reflects the subspace of high orders.
Moreover, with radius being $74$nm, we find the whole powers including absorption, scattering, and extinction, are identical to that of the core-shell system at outer radius being $66$nm, which marked by a black star symbol.
Although their constitution of scattering components is totally different, they can provide the same energy distributions, reflecting the energy degeneracy in scattering of light.

Now, we turn to discuss the existence of the superabsorption, in which the system can absorb more energy by increasing resonant order numbers while maintaining lower scattering power, as highlights in red box of Fig. 1 \cite{superabsorber}.
We consider a core-shell nanowire system made by silicon in shell and gold in core, where the ratio of inner to outer radius is fixed to $0.42$ and the outer radius is chosen as $132.98$nm.
The material dispersions we consider here are all based on  experiments, \cite{handbook}.
In Fig. 3 (a), we study energy distribution of the core-shell system with respect to varying operation wavelength from $520$nm to $540$nm, as denoted by cyan color line.
To understand its inherent scattering components, we also plot the scattering coefficients for each orders from $n=[-2,-1,0,1,2]$, as shown in Fig. 3 (b).
In this wavelength range, the dominant orders are $n=[-1,0,1]$.
We note that only at $530$nm, the cyan line would intersect with the boundary $N=1$ marked by a black star, revealing a system with lower scattering but larger absorption.
Here the normalized absorption and normalized scattering power are $[0.6,0.2]$ , respectively.
From the right side of Fig. 3 (b), it is interesting to see that the phases from dominant orders at $530$nm would be $\pi$, satisfying the superabsorption requirement.

Furthermore, we consider another system configuration by silicon in shell and silver in core with an outer radius of $51.6$nm and the ratio of inner to outer radii being $0.89$.
Fig. 3 (c) shows a brown color trajectory within the wavelength of $430$nm to $450$nm.
Here the system obviously resides in $N=0$ domain.
We mark an operating $440$nm by a black cross, corresponding to an intersection at $N=0$ boundary, with lower scattering as well as larger absorption.
The normalized absorption and normalized scattering power in this case are $[0.18,0.05]$ , respectively.
However, when analyzing the components of the scattering coefficients in Fig. 3 (d), we find the primary orders to be  $n=0$ and $n=1$.
As for its phase analysis in Fig. 3 (d), the phases of primary orders are not needed to be $\pi$.
We note that this system certainly possesses the desirable energy distribution to superabsoption, but its scattering coefficients have not met in the work \cite{superabsorber}, because this system belongs to the high scattering space domain.
This outcome reveals that when designing a subwavelengthly system associated with energy issue, it can have an opportunity to relax the constraints on inherent scattering states when considering a system with high scattering domain.

In the superscattering case, the systems can have multiple partial wave resonances occurred at same wavelength, beyond single order limit.
In appendix A, we consider an alternative system to support the similar result, but without the need of phase matching on scattering.
Before conclusion, we want to remark that although our discussion is limited to cylindrical systems, but there has a similar result to a spherical system where supports a degenerate energy characteristic.
The finding of the general energy diagram, irrespective of inherent system configurations, material choices and operating wavelength, would provide complete information in designing functional energy systems.
Last but not least, our demonstration of quasi-superabsorption and quasi-superscattering systems, without relying on resonant mechanism, would allow the system with larger quality factor.

In conclusion, with the approach of differential calculus as well as the use of Lagrange multiplier, we develop the general energy digram involving scattering, absorption, and extinction, for any passive systems.
We observe that at the boundary of energy diagram of the scatterer system, it requires scattering coefficients the same for each orders.
However, inside the diagram, there has no such correspondence to scattering coefficients.
When a system consists of higher resonant orders, the corresponding diagram can completely cover a system with lower ones.
This result reflects a energy degenerate property in light scattering.
As a result, system could have the same energy performance, but could possess different scattering states intrinsically.
We discuss systems with quasi-superabsorption and quasi-superscattering, but without the phase-matching constraints.
This work not only provides the complete information for energy distribution in the process of light scattering, but also highlights degrees of freedom in practical designs.

\section{Acknowledgement}
The author thanks Professor Ray-Kuang Lee for helpful discussions.
This work was supported by Ministry of Science and Technology, Taiwan (MOST) ($107$-$2112$-M-$259$ -$007$ -MY3).


\begin{thebibliography}{99}
\bibitem{note2}
Cross section $C$ is defined by power over intensity of incident wave and its unit is wavelength. Thus, the power $P$ we used here is the multiplication of the cross section with incident intensity.
\end{thebibliography}
\end{document}